\documentclass[10pt, conference, letterpaper]{IEEEtran}
\usepackage{graphicx}
\usepackage{epstopdf}
\usepackage{cite}
\usepackage{amsfonts}
\usepackage{amssymb}
\usepackage{amsmath}
\usepackage{mathrsfs}
\usepackage{bm}
\usepackage{amsmath}
\usepackage{mdwmath}
\usepackage{mdwtab}
\usepackage{url}
\usepackage[ruled,vlined,linesnumbered]{algorithm2e} 
\usepackage{subfigure}
\allowdisplaybreaks

\DeclareMathOperator*{\argmin}{arg\,min}
\newcommand\smallO{
	\mathchoice
	{{\scriptstyle\mathcal{O}}}
	{{\scriptstyle\mathcal{O}}}
	{{\scriptscriptstyle\mathcal{O}}}
	{\scalebox{.7}{$\scriptscriptstyle\mathcal{O}$}}
}
\newtheorem{remark}{Remark}
\newtheorem{theorem}{Theorem}
\newtheorem{lemma}{Lemma}
\newtheorem{prop}{Proposition}
\newtheorem{defi}{Definition}
\newtheorem{coro}{Corollary}

\def\H{{\boldsymbol{H}}}

\def\J{{\boldsymbol{J}}}

\def\Q{{\boldsymbol{Q}}}

\def\S{{\boldsymbol{S}}}

\def\X{{\boldsymbol{X}}}

\def\h{{\boldsymbol{h}}}

\def\r{{\boldsymbol{r}}}

\def\v{{\boldsymbol{v}}}

\def\x{{\boldsymbol{x}}}

\newenvironment{iarray}{\begin{IEEEeqnarray}{rCl}}{\end{IEEEeqnarray}\ignorespacesafterend}
\newcommand{\dx}{{\sf d}}
\usepackage{makecell,multirow,diagbox}

\begin{document}

\title{Analyzing Age of Information in Multiaccess Networks by Fluid Limits
}
\author{\IEEEauthorblockN{Zhiyuan Jiang}
	\IEEEauthorblockA{Shanghai Institute for Advanced Communication and Data Science, Shanghai University, Shanghai 200444, China.\\
		Email: jiangzhiyuan@shu.edu.cn
	}
}
\maketitle

\begin{abstract}
In this paper, we adopt the fluid limits to analyze Age of Information (AoI) in a wireless multiaccess network with many users. We consider the case wherein users have heterogeneous i.i.d. channel conditions and the statuses are generate-at-will. Convergence of the AoI occupancy measure to the fluid limit, represented by a Partial Derivative Equation (PDE), is proved within an approximation error inversely proportional to the number of users. Global convergence to the equilibrium of the PDE, i.e., stationary AoI distribution, is also proved. Based on this framework, it is shown that an existing AoI lower bound in the literature is in fact asymptotically tight, and a simple threshold policy, with the thresholds explicitly derived, achieves the optimum asymptotically. The proposed threshold-based policy is also much easier to decentralize than the widely-known index-based policies which require comparing user indices. To showcase the usability of the framework, we also use it to analyze the average non-linear AoI functions (with power and logarithm forms) in wireless networks. Again, explicit optimal threshold-based policies are derived, and average age functions proven. Simulation results show that even when the number of users is limited, e.g., $10$, the proposed policy and analysis are still effective.
\end{abstract}

\section{Introduction}
\label{sec_intro}
Age of Information (AoI) \cite{kaul12} is defined as a timeliness metric to quantify the time elapsed since the generation of information, when it is the freshest, to any specific time instant. In a time-critical networked system, it is desirable to maintain a low level of AoI at a node that depends heavily on the specific information. Therefore, the minimization of AoI is a systematic task, encompassing lowering the communication delay, sensing delay, data processing delay and considering their interplay. In this paper, we focus on the part of AoI that is attributed to wireless network scheduling, especially scenarios with a large number of users.

Many works have been dedicated specifically to solving AoI in wireless multiaccess networks. Considering active sources, i.e., the transmitted information is always fresh, the AoI is in fact identical with time-since-last-service \cite{li15}. With users having heterogeneous i.i.d. transmission successful probabilities, denoted by $q_{n}$ $\forall n\in[1,\cdots,N]$, and one of them is scheduled each time, Kadota et al. \cite{kadota18} found that the time-average AoI is lower bounded by
\begin{equation}
\label{lb1}
\bar{\Delta}_{\boldsymbol{\pi}} \ge \frac{1}{2N} \left( \sum_{n=1}^N \frac{1}{\sqrt{q_{n}}}\right)^2.
\end{equation}
The authors subsequently propose several optimization algorithms. The age-greedy policy is shown to be optimal in homogeneous networks when $q_{n}=q$, $\forall n$. The stationary randomized policy with optimal randomization is shown to be within $2$-times the optimum. Neither greedy and stationary policy achieves as good performance as the Whittle's index policy and Lyapunov-based max-weight policy, which are essentially identical regarding their scheduler criterion---they achieve very close-to-optimal performance in simulations, however can only be proven with loose bounds due to their non-renewal policy nature. Several other works also derive the Whittle's index policy in different scenarios that show similarly good performance, and again, without theoretical optimality guarantees \cite{jiang18_itc,hsu18}. A recent work by Maatouk et al. \cite{ali20} proved that Whittle's index policy is asymptotically optimal, when the number of scheduled users scales with the total number of users and they both go to infinity. Talak et al. showed that the stationary randomized policy is peak-AoI optimal even in a general network topology, wherein peak-AoI denotes the AoI right before delivery \cite{talak_mobihoc}.

Summarizing existing works, three key problems still remain to be solved. \textbf{P1}: Is the lower bound in \eqref{lb1} tight or at least asymptotically tight? What policy can be proved to achieve the optimum? \textbf{P2}: Most works focus on AoI scheduling policy design but analytical results are less available. How to analyze the AoI performance of scheduling policies? \textbf{P3}: Considering practical signaling concerns, how to decentralize the scheduler operation? 

In this paper, we address these three problems by adopting the fluid limits analysis tool. Our contributions include:

\begin{itemize}
	\item 
	We prove that the AoI occupancy measure of a general type of threshold-based policies with a large number of users follows a fluid limit. Convergence to the fluid limit, which is represented by a Partial Derivative Equation (PDE), is proved. Global stable equilibrium point of the PDE is proved, which can be used to characterize the stationary distribution of AoI.
	\item
	Based on this mathematical foundation, we show that the lower bound in \eqref{lb1} is in fact asymptotically tight, and a simple threshold-based policy achieves the optimum, with the thresholds explicitly calculated by the fluid limits. Connections to the Whittle's index policy are drawn. Analytical AoI probability distribution function by the threshold-based policy is derived, based on which AoI statistics such as moments and tail distribution can be obtained immediately.
	\item
	To showcase its usability, the framework is also applied to analyze the non-linear age function performance in wireless multiaccess networks. We derive the optimal threshold-based policy when optimizing a power or logarithm function of AoI, and the analytical AoI distribution is also obtained. Simulation results are given for better visibility.
\end{itemize}

The rest of the paper is organized as follows. In Section \ref{sec_exp}, an illustration example is given to highlight the idea of fluid limits. Section \ref{sec_model} introduces the system model and modeling language. Section \ref{sec_mf} lays down the mathematical tool that is used. Section \ref{sec_app1} and \ref{sec_app2} use the tool to analyze two important cases. Finally, in Section \ref{sec_cl}, conclusions are drawn with future work discussed.
\begin{figure}[!t]
	\centering
	\includegraphics[width=0.45\textwidth]{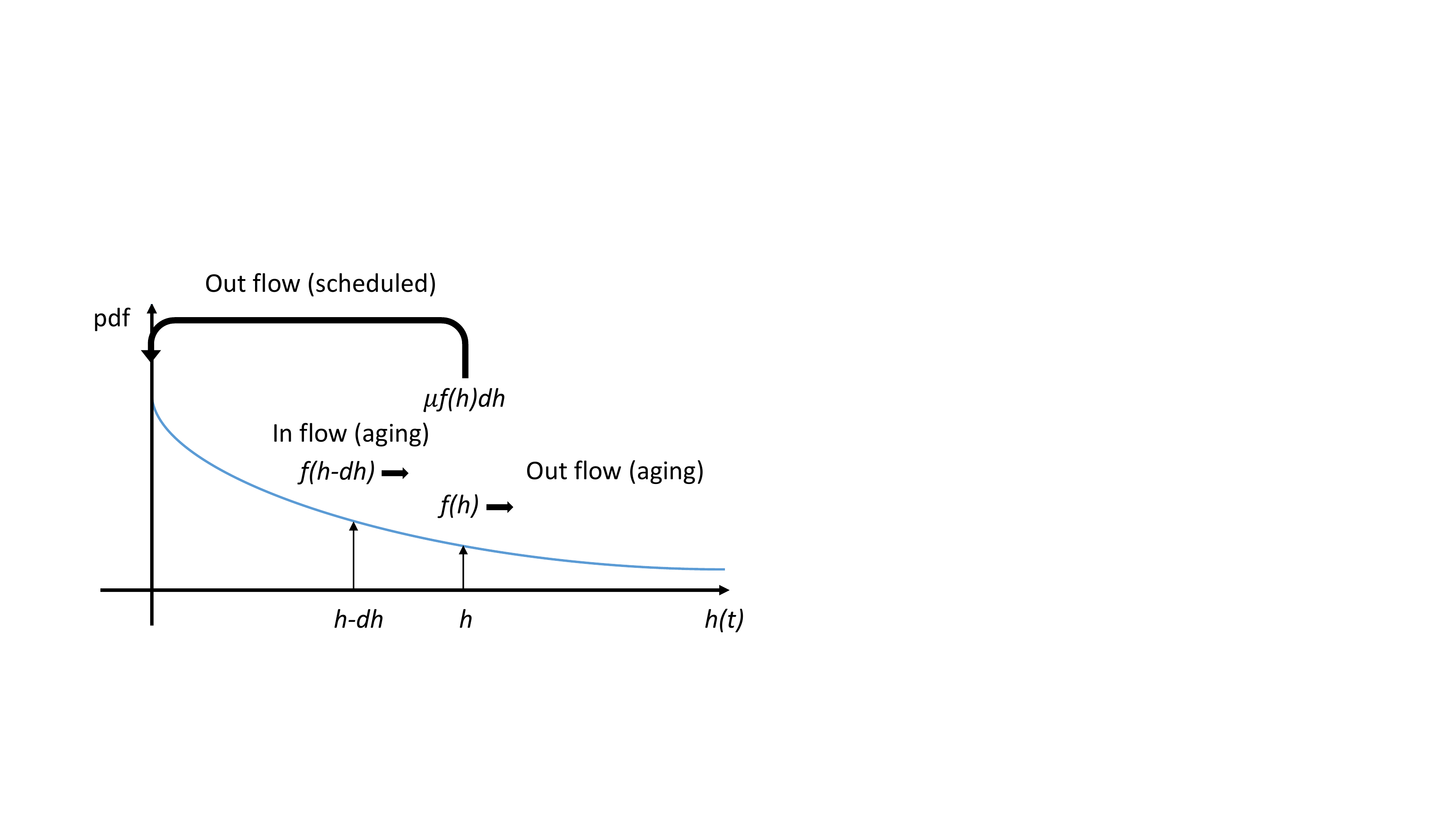}
	\caption{An exemplary AoI flow diagram assuming the fluid limit exists.}
	\label{Fig_exp}
\end{figure}
\section{Illustrating Example of AoI Fluid Limits}
\label{sec_exp}
Consider a network of $N$ users where each user is scheduled with a constant rate of $\mu$, and the scheduling events happen based on a Poisson process in continuous time. The transmission is reliable. Fig. \ref{Fig_exp} shows an exemplary occupancy measure of AoI at a certain time $t$, which is assumed to admit a probability distribution function. Assume that the system is in equilibrium and let us consider the balance equation of age flows, the left-hand side of which represents the expected in-flow to a specific AoI $h$ during a time duration of $\dx h$, and the right-hand side the expected out-flow from $h$:
\begin{equation}
\label{flow}
f(h-\dx h) = f(h)+\mu f(h) \dx h.
\end{equation}
The in-flow denotes the aging from $h-\dx h$, and the first term of the out-flow is due to aging from $h$. The second term of out-flow represents the \emph{expected} amount of agents that are scheduled and returns to AoI of zero. This formulation resembles fluids flow in continuous time and the dynamic is characterized by an Ordinary Differential Equation (ODE). When $\dx h$ is sufficiently small, we obtain
\begin{equation}
f^\prime = -\mu f,\, f(0)=\mu.
\end{equation}
The initial condition is obtained by considering the flow at $h=0$. This yields a solution which reads
\begin{equation}
f=\mu e^{-\mu h},\,h>0.
\end{equation}

Alternatively, one can solve for the solution of an embedded Markov chain describing the evolution of AoI and obtain the exactly same AoI stationary distribution, which is omitted here due to brevity. The rationality of this fluid approach is that, with growing number of agents, one expects that the law of large numbers takes effect, such that the expected flow in \eqref{flow}, which takes expectation in the probability sense, coincides with the actual fraction of agents that are scheduled. This is called the fluid limits. Consequently, the calculation of AoI statistics follows immediately.

Three key questions exist in applying this idea. \textbf{Q1}: What are the conditions allowing the limiting ODE or PDE (partial DE), as well as its equilibrium, to exist? \textbf{Q2}: In this example, the expected scheduled agents per time unit is $\mu N$ which also goes to infinity. In practice, it is more reasonable to consider a finite number of scheduled agents. Does the fluid limit also exist? \textbf{Q3}: Can we use this to solve particular problems in AoI? In following sections, we endeavor to answer these questions.
 
\section{Fluid Limit Modeling Language for AoI}
\label{sec_model}
In this section, fluid limit modeling specifications for AoI evolution of many users (hereinafter we use agents) are introduced, which are described in a general setting such that they can be adopted in as many scenarios as possible. We first describe a Continuous Time Markov Chain (CTMC)-based formulation, which is also known as the Markov processes of pure jump type. However, this formulation is problematic in considering the convergence of the AoI occupancy measure. Therefore, it is transformed to a rescaled time domain. Connections to the widely-used Discrete Time Markov Chain (DTMC)-based AoI model \cite{kadota18,jiang18_iot}, which is inconvenient to apply the fluid limits, are also discussed.

\subsection{Multi-Class CTMC-Based AoI Model}
We assume that agents in the system belong to a finite number of classes, which are prescribed based on their features, e.g., channel conditions, packet arrival patterns, performance metrics. Agents within each class are considered exactly the same and hence interchangeable, i.e., agents having identical state evolution dynamics. In practice, this assumption can be relaxed to be approximately the same. The total number of agents and classes are denoted by $N$ and $C$ respectively, with $N_c$ denotes the total number of agents in class-$c$, hence $\sum_{c=1}^C N_c=N$. The evolution of the system is captured by a CTMC, which is described as follows.

\begin{defi}
	\label{def1}
	A CTMC model for AoI is a tuple $\mathcal{H}=(\boldsymbol{H},\mathcal{T},\boldsymbol{d}_0)$, where:
	\begin{enumerate}
		\item 
		$\boldsymbol{H}=(\boldsymbol{h}_1,...,\boldsymbol{h}_C)$ denotes the vector of AoI of $N$ terminals categorized by $C$ classes, wherein $\boldsymbol{h}_c=(h_{c,1},\cdots,h_{c,N_c})$. Each $h_{c,i}$  takes value in $\mathbb{R}_{\ge 0}$. $\boldsymbol{d}_0 \in \mathbb{R}_{\ge 0}$ is the initial state of the model.
	    \item
	   	$\mathcal{T}=\{\tau_1,\cdots,\tau_M\}$ denotes the set of transitions of the form $\tau_j=(a,\phi(\boldsymbol{H}),\boldsymbol{v}(\boldsymbol{H}),r(\boldsymbol{H}))$, where:
	   	\begin{enumerate}
	   		\item 
	   		The label of the transition is denoted by $a$.
	   		\item
	   		$\phi(\boldsymbol{H})$ denotes the set of inequalities that need to be satisfied by the transition, indicating the conditions that such a transition happens.
	   		\item
	   		The update vector is denoted by $\boldsymbol{v}(\boldsymbol{H}) \in \mathbb{R}^N$, representing the net change on the state by the transition. It is required that $\boldsymbol{H} + \v(\boldsymbol{H}) \in \mathbb{R}_{\ge 0}$ whenever $\phi(\boldsymbol{H})$ is true.
	   		\item
	   		The transition rate function is denoted by $r(\boldsymbol{H}): \mathbb{R}_{\ge 0} \to \mathbb{R}_{\ge 0}$, which specifies the transition as a function of the current state and is required to be Lipschitz continuous and bounded in general. The time interval between successive transitions is assumed to be exponentially distributed with rate $r(\boldsymbol{H})$. 
	   	\end{enumerate}
	\end{enumerate}

This definition is described in a quite general way. To give a concrete example, let us consider an AoI scheduling problem based on the Whittle's index \cite{jiang18_itc,hsu18}, wherein the index for agent-$n$ at time-$t$ is $w_n(t)$ and the agent with the highest index is scheduled at every time slot. The transmission is assumed to be successful with probability $p_\mathsf{s} \in (0,1]$ due to channel error, and in the case of success, the AoI decreases to zero. The mean time interval between successive scheduling events is set to $1/r_0$. The corresponding transitions can be described as (using agent-$1$ as a representative):
\begin{enumerate}
	\item 
	$\tau_1=(\textrm{Scheduled and successful update},w_1(t)>w_{i}(t)(\forall i\neq 1),(-h_{1,1}(t),\Delta(t),\cdots,\Delta(t)),p_\mathsf{s} r_0)$;
	\item 
	$\tau_2=(\textrm{Scheduled and unsuccessful update},w_1(t)>w_{i}(t)(\forall i\neq 1),(\Delta(t),\cdots,\Delta(t)),(1-p_\mathsf{s})r_0)$,
\end{enumerate}
wherein $\Delta(t)$ denotes the time elapsed since the last scheduling event. 
\end{defi}

Define the occupancy measure among all agents, with AoI of $\H(t)$ in class-$c$ by
\begin{equation}
	\label{x}
X_{c,t}^{N}(h) = \frac{1}{N} \sum_{i=1}^{N_c} \delta_{h_{c,i}(t)}(h),
\end{equation}
where $\delta_y(x)$ denotes the Dirac delta distribution which is a linear mapping: $D(\mathbb{R})\rightarrow\mathbb{R}$ that satisfies $\int \delta_y(x) f(x) \dx x = f(y)$ and $D(\mathbb{R})$ denotes the set of test functions. Denote the faction of agents that are in class $c$ and with AoI lower than $h$ as 
\begin{equation}
F_{c,t}^N(h) = \int_0^h X_{c,t}^{N}(x) \dx x
\end{equation}
\begin{remark}
	One critical issue of this formulation, making it inapplicable for the fluid limits, is that when the system size grows large, i.e., $N$ is large, the resultant occupancy measure of agents' AoI is not tight (definition of tightness of measure is given in \eqref{tightness}), in fact it has support approaching infinity, and therefore not convergent to any probability measure. This shall be made evident in the proof of the main theorem. To address this, a time-rescaled CTMC formulation is introduced as follows.
\end{remark}
\subsection{Transformation to Rescaled CTMC}
\begin{defi}
	A time-rescaled CTMC model for AoI is a tuple $\hat{ \mathcal{H}}=(\hat{\boldsymbol{H}}, \hat{\mathcal{T}},\hat{\boldsymbol{d}_0)}$, which is defined identically with Definition \ref{def1}, except that $\hat r(\boldsymbol{H}) = N r(\boldsymbol{H})$, $\forall \boldsymbol{H}$.
\end{defi}

In essence, $\hat{ \mathcal{H}}$ can be viewed as an accelerated version of ${ \mathcal{H}}$, in the sense that time intervals between scheduling events are scaled down with the number of agents. By a stochastic coupling argument, it is clear that the following lemma is true because everything happens sooner.
\begin{lemma}
	The occupancy measure of $\hat{ \mathcal{H}}$ satisfies
	\begin{equation}
		\label{F}
	\hat F_{c,t}^{N}(h) = F_{c,t}^{N}(Nh),
	\end{equation}
	and consequently, if $\hat X_{c,t}^{N}(h)$ converges weakly to a probability measure $\hat x_{c,t}(h)$ with $N \to \infty$ (later formalized), the mean AoI of ${ \hat {\mathcal{H}}}$ (i.e., $\mathbb{E} [\hat h]$) and ${ {\mathcal{H}}}$ (i.e., $\mathbb{E} [ h]$) satisfy 
	\begin{equation}
	\label{meanAoI}
	\mathbb{E} [\hat h] = \sum_{c=1}^C\int_0^{\infty} h \hat x_{c,t}(h) \dx h=\frac{\mathbb{E}[h]}{N}
	\end{equation}
	
\end{lemma}

Fortunately, by this definition, we can prove in many practical cases the occupancy measure $\hat X_{c,t}^{N}(h)$ is tight in the fluid limit regime, and hence converges weakly to a continuous probability measure, which facilitates our subsequent analysis. 

On the flip side, by making this time rescaling, we are unable to analyze the AoI exactly. As we shall see in the next section, by adopting the fluid limit approximation, the approximation error is $\mathcal{O}(\frac{1}{N})$. Combined with \eqref{meanAoI}, it is clear that we can only solve for the linear scaling factor with $N$, i.e., asymptotic optimality---which is sufficient in many cases because AoI indeed scales linearly with $N$, and when $N$ gets large, the error becomes infinitesimal. 
\subsection{Connection with DTMC Formulation}
In most previous works on AoI scheduling, e.g., \cite{kadota18,jiang18_iot,hsu18,ali20}, the AoI evolution is modeled by a DTMC. In this paper, we adopt CTMC, whereas it is desired to draw the connection between them. In fact, the DTMC can be viewed as the embedded DTMC of the CTMC. More specifically, denote the infinitesimal generator matrix $\Q$ of $\mathcal{H}$ as
\begin{equation}
q_{\H,\H^\prime}=\sum\{\r(\boldsymbol{H})|\tau \in \mathcal{T},\phi{(\X)}\textrm{ is true},\H^\prime = \H+\v(\H)\},
\end{equation}
and set the transition rate out of each state to one, i.e., the mean scheduling interval is $1$, that is
\begin{equation}
r_{-\H}=\sum_{\H^\prime \neq \H} q_{\H,\H^\prime}=1.
\end{equation}
Leveraging the rescaling technique, we obtain $\hat{\mathcal{H}}$. The total number of update attempts (i.e., scheduling events) with rescaled CTMC is hence $\mathcal{N}(tN)$ where $\mathcal{N}$ is a Poisson random variable with mean rate of $r_{-\H}$, and the number is $\left\lfloor tN \right\rfloor$ for the DTMC. We obtain
\begin{equation}
\lim_{N \to \infty} \frac{\mathcal{N}(tN)}{\left\lfloor tN \right\rfloor} = \lim_{N \to \infty} \frac{\mathcal{N}(tN)/tN}{\left\lfloor tN \right\rfloor/tN}=1,
\end{equation}
which is attributed to the law of large numbers. In other words, this equation shows that by adjusting the rate functions of CTMC, with $N$ growing large, the same number of scheduling events have happened for both CTMC and DTMC at any time $t$, making them statistically equivalent. Using the uniformization technique \cite{gal12} can also leads to this conclusion.
\section{Deterministic Fluid Limit of AoI and Stationary Regime}
\label{sec_mf}
In this section, we develop our main result on the fluid limit of AoI based on the time-rescaled CTMC model. We consider a specific but widely-adopted type of scheduling policies for AoI, which is \emph{threshold-type} policies. In particular, we assume
\begin{enumerate}
	\item
	At every scheduling event, one agent\footnote{Generalization to a \emph{finite} number of agents is straightforward. However, an infinite number of scheduled agents \cite{ali20,whitt84} needs a different formulation.} is randomly chosen from the agents with AoI greater than their (possibly different) thresholds with equal probabilities. In the case of no agent has AoI above its threshold, the system stays idle.
	\item 
	Agents in each class have the same threshold $H_c$.
\end{enumerate}

Note that the threshold-type policy is adopted by many papers \cite{ceran19,tang20,jiang19_tcom,ali20}. It is both intuitive, in the sense that AoI higher than a threshold should be prioritized, and proven optimal in many scenarios. However, usually the exact thresholds cannot be solved explicitly. In fact, we will show that  even in scenarios wherein threshold-type polices are not considered optimal, and index-based policies \cite{kadota18,jiang18_itc,hsu18} have superior performance, an optimized threshold-type policy can achieve asymptotically optimal AoI with proven and closed-form performance expressions.
\subsection{Fluid Limit}
When $N$ gets large, the evolution of the time-rescaled CTMC becomes close to a deterministic fluid limit, characterized by differential equations. 
\begin{theorem}
	\label{thm_mf}
	Assume that the initial occupancy measure $\hat{\X}_{0}^{N}$ converges weakly to a deterministic distribution $\hat{\x}_{0}$ which admits a density $\{\hat f_{c,0}|c=1,\cdots,C\}$. Then as $N$ approaches infinity, $\hat{X}_{c,t}^{N}$ of each class converges in distribution to a deterministic fluid limit processes $\{\hat{x}_{c,t}|t \in \mathbb{R}_{\ge 0}\}$ which admits a density for all $t$ denoted by $\hat f_{c}(t,h)$ and CDF by $\hat F_{c}(t,h)$, is the unique solution of the following PDE:
	\begin{iarray}
		\label{pde}
		&& \frac{\partial{\hat f_{c}(t,h)}}{\partial{t}}  =\nonumber\\ && \left\{\,
		\begin{IEEEeqnarraybox}[][c]{l?l}
			\IEEEstrut	
			-\frac{\partial{\hat f_{c}(t,h)}}{\partial{h}} - \frac{p_{\mathsf{s},c}\hat f_{c}(t,h)}{1-\sum_{c=1}^C \hat F_{j}(t,\hat H_j)}, &\textrm{if $h>\hat H_c$} ; \\
			-\frac{\partial{\hat f_{c}(t,h)}}{\partial{h}} ,&\textrm{if $0 \le h \le \hat H_c$},
			\IEEEstrut
		\end{IEEEeqnarraybox}
		\right. 	\\	
	  && \forall h\ge 0, \forall c, \,\hat f_{c}(0,h) = \hat f_{c,0},\\
	  && \forall t\ge0,\forall c, \,\hat f_{c}(t,0) = \frac{\eta_c- \hat F_{c}(t,\hat H_c)}{1-\sum_{j=1}^C \hat F_{j}(t,\hat H_j)}p_{\mathsf{s},c},
	  \label{t2}
	\end{iarray}
	where $\eta_c = \frac{N_c}{N}$.
\end{theorem}
\begin{IEEEproof}
The proof is based on previous works on fluid limits and mean-field approximations \cite{eth09,mftt13,kol12,gossip09,bord07}. First, define the characteristic function, i.e., $\varphi^N_{\hat H,c}(\omega,t)$: $\mathbb{R} \times \mathbb{R}_{\ge 0} \rightarrow \mathbb{C}$, of the occupancy measure as 
\begin{equation}
	\varphi^N_{\hat H,c}(\omega,t) \triangleq \int_0^\infty e^{j\omega h} \dx \hat F_{c,t}^{N}(h)=\int_0^\infty e^{j\omega h} \hat X_{c,t}^{N}(h) \dx h.
\end{equation}
For ease of notation, we use $\varphi_c^N(t)$ for $\varphi^N_{\hat H,c}(\omega,t)$. Note that by definition and \eqref{x}-\eqref{F}, $\varphi_c^N(t) = \frac{1}{N} \sum_{i=1}^{N_c} e^{j\omega h_{c,i}(t)}$, which is the sample average. Consider the generator of $\varphi_c^N(t)$:
\begin{equation}
	\mathcal{G}(\varphi_c^N(t)) \triangleq \lim_{\dx t \rightarrow 0} \frac{\mathbb{E}[\varphi_c^N(t+\dx t)-\varphi_c^N(t)|\mathcal{F}_t]}{\dx t},
\end{equation}
where $\mathcal{F}_t$ is the natural filtration of $\hat X_{c,t}^{N}(h)$. 
\begin{lemma}
	Define 
	\begin{equation}
		M^N_{t,c} \triangleq \varphi_c^N(t) - \varphi_c^N(0) - \int_0^{t} \mathcal{G}(\varphi_c^N(s)) \dx s,
	\end{equation}
then $M^N_t$ is a zero-mean $\mathcal{F}_t$-martingale.
\end{lemma}
\begin{IEEEproof}
	For any $t>s$,
	\begin{iarray}
		&& \mathbb{E}[M^N_{t,c}|\mathcal{F}_s] \nonumber\\
		&=& M^N_{s,c} + \mathbb{E}\left[\varphi_c^N(t)-\varphi^N(s)-\left.\int_s^t \mathcal{G}(\varphi_c^N(s)) \dx s\right|\mathcal{F}_s\right] \nonumber\\
		&=& M^N_s + \mathbb{E}\left[\varphi_c^N(t)-\varphi_c^N(s)|\mathcal{F}_s\right]-\left(\mathbb{E}\left[\varphi_c^N(t)|\mathcal{F}_s\right]-\varphi_c^N(s)\right)\nonumber\\
		&=& M^N_s.
	\end{iarray}
The mean follows directly from the definition.
\end{IEEEproof}

Then we calculate $\mathcal{G}(\varphi^N(t))$ specifically as:

\begin{equation}
	\mathcal{G}(\varphi_c^N(t)) = G_{c,1} + G_{c,2},
\end{equation}
\begin{iarray}
	G_{c,1} &=& \lim_{\dx t \rightarrow 0} \frac{1}{\dx t} \left[\int_0^\infty e^{j\omega h} \frac{1}{N} \sum_{i=1}^{N_c} \delta_{h_{c,i}(t)+\dx t} \dx h \right. \nonumber\\
	&& \left. - \int_0^\infty e^{j\omega h} \frac{1}{N} \sum_{i=1}^{N_c} \delta_{h_{c,i}(t)} \dx h\right],\nonumber\\
	&=& \lim_{\dx t \rightarrow 0} \frac{1}{\dx t} \left[ \frac{1}{N} \sum_{i=1}^{N_c} e^{j\omega ({h_{c,i}(t)+\dx t})} -\frac{1}{N} \sum_{i=1}^{N_c} e^{j\omega ({h_{c,i}(t)})} \right], \nonumber\\
	&=& j\omega \frac{1}{N} \sum_{i=1}^{N_c} e^{j\omega {h_{c,i}(t)}} \nonumber\\
	G_{c,2} &=& \lim_{\dx t \rightarrow 0} \frac{p_{\mathsf{s},c}}{\beta_t N \dx t} \mathbb{E}[\Lambda(N,\dx t)] \nonumber\\
	&& \cdot \sum_{\substack{j=1,\\h_{c,j}(t)>H_c}}^{N_c} \left[\int_0^\infty e^{j\omega h} \frac{1}{N} \left(\sum_{i=1}^{N_c} \delta_{h_{c,i}(t)} + \delta_0 \right.\right.\nonumber\\
	&& \left.\left. - \delta_{h_{c,j}(t)} \right) \dx h - \int_0^\infty e^{j\omega h} \frac{1}{N} \sum_{i=1}^{N_c} \delta_{h_{c,i}(t)} \dx h\right],\nonumber\\
	\label{21}
	&=& \frac{p_{\mathsf{s},c}}{\beta_t}   \int_0^\infty e^{j\omega h} \sum_{\substack{j=1,\\h_{c,j}(t)>H_c}}^{N_c} \frac{1}{N} \left( \delta_0  - \delta_{h_{c,j}(t)} \right) \dx h, \\
	\beta_t &=& \int_0^\infty \sum_{\{c, i|h_{c,i}(t) > H_c\} } \delta_{h_{c,i}(t)} \dx h = \sum_{c=1}^C \frac{N_{h_t> H_c}}{N},
\end{iarray}
wherein $\Lambda(N,\dx t)$ denotes a Poisson distributed random variable with rate $N$ and time of $\dx t$, and $N_{h_t> H_c}$ denotes the number of agents in class $c$ that have AoI larger than the threshold. The last equality in \eqref{21} is because when $\dx t \to 0$, the Poisson distributed $\Lambda(N,\dx t)$ is closely approximated by a Bernoulli random variable with parameter $N\dx t$. The calculation above derives the expected infinitesimal change in $\varphi_c^N(t)$. Specifically, with probability $\frac{p_{\mathsf{s},c}}{\beta_t N \dx t}$ which represents that an agent is chosen uniformly randomly from all eligible agents with AoI larger than their thresholds, an agent is scheduled and updates successfully with AoI returning to zero---such an event happens with a mean rate of $N$ in the time-rescaled CTMC. 

Now let us consider the quadratic variation of $M^N_{t,c}$. The generator of $\|\varphi_c^N(t)\|^2$ is 
\begin{equation}
	\label{gsquare}
	\mathcal{G}(\|\varphi_c^N(t)\|^2) = G^\prime_{c,1} + G^\prime_{c,2},
\end{equation}
\begin{iarray}
	G^\prime_{c,1} &=& \lim_{\dx t \rightarrow 0} \frac{1}{\dx t} \left[\left(\int_0^\infty e^{j\omega h} \frac{1}{N} \sum_{i=1}^{N_c} \delta_{h_{c,i}(t)+\dx t} \dx h\right)^2 \right. \nonumber\\
	&& \left. - \left( \int_0^\infty e^{j\omega h} \frac{1}{N} \sum_{i=1}^{N_c} \delta_{h_{c,i}(t)} \dx h\right)^2\right],\nonumber\\
	&=& 2j\omega  \left(\frac{1}{N}\sum_{i=1}^{N_c} e^{j\omega {h_{c,i}(t)}}\right)^2 \nonumber\\
	G^\prime_{c,2} &=& \lim_{\dx t \rightarrow 0} \frac{p_{\mathsf{s},c}}{\beta_t N \dx t} \mathbb{E}[\Lambda(N,\dx t)] \nonumber\\
	&& \cdot \sum_{\substack{j=1,\\h_{c,j}(t)>H_c}}^{N_c} \left[\left(\int_0^\infty e^{j\omega h} \frac{1}{N} \left(\sum_{i=1}^{N_c} \delta_{h_{c,i}(t)} + \delta_0 \right.\right.\right.\nonumber\\
	&& \left. \left.\left. - \delta_{h_{c,j}(t)} \right) \dx h\right)^2 -\left( \int_0^\infty e^{j\omega h} \frac{1}{N} \sum_{i=1}^{N_c} \delta_{h_{c,i}(t)} \dx h \right)^2\right],\nonumber\\
	&=& \frac{2p_{\mathsf{s},c}}{\beta_t}    \frac{1}{N}  \sum_{\substack{j=1,\\h_{c,j}(t)>H_c}}^{N_c} \left(1-e^{j\omega h_{c,j}(t)}\right)  \frac{1}{N} \sum_{i=1}^{N_c} e^{j\omega {h_{c,i}(t)}} \nonumber\\
	\label{23}
	&& + \frac{p_{\mathsf{s},c}}{\beta_t}    \frac{1}{N^2}  \sum_{\substack{j=1,\\h_{c,j}(t)>H_c}}^{N_c} \left(1-e^{j\omega h_{c,j}(t)}\right)^2. 
\end{iarray}
Note that we obtain the second term in \eqref{23} as
\begin{equation}
	\frac{p_{\mathsf{s},c}}{\beta_t}    \frac{1}{N^2}  \sum_{\substack{j=1,\\h_{c,j}(t)>H_c}}^{N_c} \left(1-e^{j\omega h_{c,j}(t)}\right)^2 \le \frac{4 p_{\mathsf{s},c} N_{h_t> H_c}}{\beta_t N^2} \le  \frac{4 p_{\mathsf{s},c} }{N},   
\end{equation}
where the last inequality follows from $\beta_t > \frac{N_{h_t> H_c}}{N}$, $\forall c$. The term goes to zero with $N$ getting large, i.e., define 
\begin{equation}
	\bar{G}^\prime_{c,2}=\frac{2p_{\mathsf{s},c}}{\beta_t}    \frac{1}{N}  \sum_{\substack{j=1,\\h_{c,j}(t)>H_c}}^{N_c} \left(1-e^{j\omega h_{c,j}(t)}\right)  \frac{1}{N} \sum_{i=1}^{N_c} e^{j\omega {h_{c,i}(t)}}.
\end{equation}
 It follows that 
\begin{equation}
	\label{msquare}
	M^{\prime N}_{t,c} \triangleq \|\varphi_c^N(t)\|^2 - \|\varphi_c^N(0)\|^2 - \int_0^{t} \mathcal{G}(\|\varphi_c^N(s)\|^2) \dx s
\end{equation}
is also a $\mathcal{F}_t$-martingale. The following lemma establishes the limiting behavior of the occupancy measure $X_{c,t}^{N}(h)$.

\begin{lemma}
	When $N$ goes to infinity, any sequence $X_{c,t}^{N}(h)$ has a convergent subsequence whose limit is denoted by $\hat f_{c,t}(h)$, and $\hat f_{c,t}(h)$ has continuous sample path.
\end{lemma}
\begin{IEEEproof}
Let us first prove the tightness of $X_{c,t}^{N}(h)$. From \cite[Lemma 4.6]{gossip09}, it is sufficient to show the tightness of AoI measure of agent-$1$, i.e., $\hat h_{c,1}(t) =h_{c,1}(t) /N$. Based on \cite[Corollary 4.3]{kurtz81}, the following needs to uphold: For every $\epsilon,T>0$, there exists $H>0$ such that
\begin{equation}
	\label{tightness}
	\inf_N \Pr\left(\hat h_{c,1}(t) \le H,\, \forall t \le T\right) > 1-\epsilon.
\end{equation} 
In the threshold-based algorithm, agent-$1$ is scheduled and the update is successful with a rate no less than $p_{\mathsf{s},c}/N$ when its AoI is larger than the threshold, i.e., $\hat h_{c,1}(t) > H_c$. Denote $T_0(t)$ as the last time agent-$1$ is scheduled and the transmission is successful. Then $\hat h_{c,1}(t) = t-T_0(t)$ is stochastic-dominated by a random variable $\hat h^\prime = H_c + \hat z$ where $\hat z \sim \textrm{exp}(p_{\mathsf{s},c})$ (in the rescaled time domain) because the scheduling event happens with a rate of $N$. Therefore,
\begin{equation}
	\label{hbound}
	\Pr\left(\hat h_{c,1}(t) \le H\right) \ge \Pr\left(\hat h^\prime \le H\right) = 1-e^{-p_{\mathsf{s},c}\left(H-H_c\right)}.
\end{equation}
Letting the right-hand side be larger than $1-\epsilon$ gives us the conclusion that $X_{c,t}^{N}(h)$ is tight. Then based on the Prokhorov's Theorem, $X_{c,t}^{N}(h)$ is relatively tight, which means that for any sequence $X_{c,t}^{N}(h)$, there exists a subsequence that is convergent. Let us denote such a subsequence by $X_{c,t}^{N_k}(h)$. Consider its characteristic function,
\begin{equation}
	\varphi_{c}(t) \triangleq \int_0^\infty e^{j\omega h} X_{c,t}^{N_k}(h) \dx h.
\end{equation}
When the time changes from $t$ to $t + \dx t$, the change in characteristic function is
\begin{iarray}
	|\varphi_{c}(t + \dx t) -\varphi_{c}(t)| &=& \int_0^\infty e^{j\omega h} \left|X_{c,t + \dx t}^{N_k}(h) -X_{c,t}^{N_k}(h)\right| \dx h.\nonumber\\
	&\le& \int_0^\infty \left|X_{c,t + \dx t}^{N_k}(h) -X_{c,t}^{N_k}(h)\right| \dx h.
\end{iarray}
Choose an arbitrary $\varepsilon > 0$. Based on \eqref{tightness}, choose $H$ so large that $\Pr\left(\hat h_{c,1}(t) > H\right) < \frac{\varepsilon}{4}$, and $N_k>\frac{4H}{\varepsilon}$, then 
\begin{iarray}
	|\varphi_{c}(t + \dx t) -\varphi_{c}(t)| &\le& \int_0^H \left|X_{c,t + \dx t}^{N_k}(h) -X_{c,t}^{N_k}(h)\right| \dx h + \frac{\varepsilon}{2}\nonumber\\
	&\le& \frac{2H}{N_k}+\frac{\varepsilon}{2}=\varepsilon.
\end{iarray}
The last inequality is based on the fact that the jump of $X_{c,t}^{N_k}(h)$ in a small time interval is at most $\frac{\delta_x}{N_k}$. This concludes the proof.
\end{IEEEproof}

Now we know that any sequence $X_{c,t}^{N}(h)$ has a convergent subsequence with continuous sample path. Let us see if such a limit, i.e., $\hat f_{c,t}(h)$, is uniquely determined. It follows from \eqref{msquare} that when $N$ is large,
\begin{equation}
	\label{32}
	 \|\varphi_c^\infty(t)\|^2= \|\varphi_c^\infty(0)\|^2 + \int_0^{t} ({G}^\prime_{c,1}+\bar{G}^\prime_{c,2}) \dx s + M^{\prime \infty}_{t,c}.
\end{equation}
Applying the Ito's formula to $f(\varphi_c^\infty(t))$ wherein $f(x)=x^2$ is twice continuously differentiable and $\varphi_c^\infty(t)$ is a continuous semimartingale, we obtain
\begin{iarray}
	\label{33}
	&& f(\varphi_c^\infty(t)) = \|\varphi_c^\infty(t)\|^2 \nonumber\\
	&=& f(\varphi_c^\infty(0)) + \int_0^{t} f^\prime(\varphi_c^\infty(s)) \dx \varphi_c^\infty(s) \nonumber\\
	&& + \frac{1}{2} \int_0^{t} f^{\prime \prime} (\varphi_c^\infty(s))  \dx \langle 	M^\infty_{s,c} \rangle \nonumber\\
	&=& \|\varphi_c^\infty(0)\|^2 + \int_0^{t} \varphi_c^\infty(s) \dx \varphi_c^\infty(s) + \langle 	M^\infty_{s,c} \rangle \nonumber\\
	&=& \|\varphi_c^\infty(0)\|^2 + 2\int_0^{t} \varphi_c^\infty(s) \mathcal{G}(\varphi_c^\infty(s)) \dx s + \langle 	M^\infty_{s,c} \rangle \nonumber\\
	&& + 2\int_0^{t} \varphi_c^\infty(s) \dx M^\infty_{s,c}\nonumber\\
		&=& \|\varphi_c^\infty(0)\|^2 + 2\int_0^{t} \varphi_c^\infty(s) (G_{c,1} + G_{c,2}) \dx s + \langle 	M^\infty_{s,c} \rangle \nonumber\\
	&& + 2\int_0^{t} \varphi_c^\infty(s) \dx M^\infty_{s,c}\nonumber\\
	&=& \|\varphi_c^\infty(0)\|^2 + \int_0^{t} ({G}^\prime_{c,1}+\bar{G}^\prime_{c,2}) \dx s + \langle 	M^\infty_{s,c} \rangle \nonumber\\
	&& + 2 \int_0^{t} \varphi_c^\infty(s) \dx M^\infty_{s,c},
\end{iarray}
wherein $ \langle X(t) \rangle \triangleq \lim_{\|P\| \to 0} \sum_{k=1}^n (X_{t_k}-X_{t_{k-1}})^2$ denotes the quadratic variation of $X(t)$ and $P \triangleq \sup_k |t_k-t_{k-1}|$. Comparing \eqref{32} and \eqref{33}, and by the uniqueness of the Doob–Meyer decomposition, we know that 
\begin{equation}
	\langle 	M^\infty_{s,c} \rangle = 0, \forall t.
\end{equation}
Based on the definition,
\begin{iarray}
	&&\mathbb{E}\left[\langle 	M^\infty_{s,c} \rangle\right] = 0 \nonumber\\
	&=& \lim_{\|P\| \to 0} \sum_{k=1}^n 2 \mathbb{E}	\|M^\infty_{t_k,c}\|^2 - 2 \mathbb{E}	\left[M^\infty_{t_{k},c}M^\infty_{t_{k-1},c}\right] \nonumber\\
	&=& \lim_{\|P\| \to 0} \sum_{k=1}^n 2 \mathbb{E}	\|M^\infty_{t_k,c}\|^2 - 2 \mathbb{E}	\|M^\infty_{t_{k-1},c}\|^2 \nonumber\\
	&=& 2 \mathbb{E}	\|M^\infty_{t_k,c}\|^2.
\end{iarray}
Therefore $M^\infty_{s,c}$ is a zero-mean martingale with zero variance, hence it is zero almost surely. Thus we obtain
\begin{equation}
	\label{intg}
\varphi_c^\infty(t) = \varphi_c^\infty(0) + \int_0^{t} (G_{c,1} + G_{c,2}) \dx s.
\end{equation}
Take the inverse characteristic function transform and derivative with $t$ on both ends, we obtain exactly the PDE in \eqref{pde}. Based on the Cauchy-Lipschitz theorem, \eqref{pde} has one unique solution. Because any solution of \eqref{intg} must also satisfy \eqref{pde}, hence $\hat f_{c,t}(h)$ is uniquely determined. Based on the following corollary of the Prokhorov's theorem, we have concluded the proof.
\begin{lemma}
	If $\mu _{n}$ is a tight sequence of probability measure such that every weakly convergent subsequence $\mu _{n_{k}}$ has the same limit $\mu$, then the sequence $\mu _{n}$ converges weakly to $\mu$.
\end{lemma}
\end{IEEEproof}

The intuition behind \eqref{pde} is clear. As the time passes, the AoI gets larger which is denoted by the first term on the right-hand side of the equation. When the AoI is below the threshold of this class, the agents are never scheduled, resulting in the bottom equation; otherwise, the agents with AoI higher than the threshold are scheduled and flow back to zero AoI with rate the proportional between the number of agents with this AoI and the total number of agents with AoI beyond their individual thresholds.

\subsection{Stationary Regime}
\label{sec_sta}
In \eqref{pde}, the AoI distribution and evolution of the system is characterized by a PDE. Let us first search for its solution. Define $	\hat g_{c}(x,y) \triangleq \hat f_{c}(x+y,x)$, wherein we use the change of variable $x=h$ and $y=t-h$. 
\begin{equation}
	\frac{\partial{\hat g_{c}(x,y) }}{\partial{x}} = \frac{\partial{\hat f_{c}(t,h)}}{\partial{t}} + \frac{\partial{\hat f_{c}(t,h)}}{\partial{h}}
\end{equation}
Therefore the PDE in \eqref{pde} is rewritten as
\begin{iarray}
	\label{pdeG}
	&& \frac{\partial{\hat g_{c}(x,y) }}{\partial{x}}  =\nonumber\\ && \left\{\,
	\begin{IEEEeqnarraybox}[][c]{l?l}
		\IEEEstrut	
		- \frac{p_{\mathsf{s},c}\hat g_{c}(x,y)}{1-\sum_{c=1}^C \hat G_{j}(\hat H_j,y)}, &\textrm{if $x>\hat H_c$} ; \\
		0 ,&\textrm{if $0 \le x \le \hat H_c$},
		\IEEEstrut
	\end{IEEEeqnarraybox}
	\right. 	\\	
	&& \forall y\le0, \forall c, \,\hat g_{c}(-y,y) = \hat f_{c,0}(-y),\\
	\label{init2}
	&& \forall y > 0,\forall c, \,\hat g_{c}(0,y) = \frac{\eta_c- \hat G_{c}(\hat H_c,y)}{1-\sum_{j=1}^C \hat G_{j}(\hat H_j,y)}p_{\mathsf{s},c},
\end{iarray}
wherein $\hat G_{c}(x,y) \triangleq \int_0^x \hat g_{c}(s,y) \dx s$. The derivative of $\hat g_{c}(s,y)$ is hence shown to be only related to $x$ and $y$ affects the initial conditions for various $x$. This helps us to derive the solution of \eqref{pde}. The following theorem describes the limiting behavior of AoI.
\begin{theorem}
	\label{thm_sta}
	With probability $1$, when $t \to \infty$, the solution of \eqref{pde} converges to a globally-stable equilibrium that is irrelevant with $t$ and $\hat f_{c,0}$, which reads
	\begin{iarray}
		\label{stas}
	 \hat d_{c}(h)  & = & \left\{\,
		\begin{IEEEeqnarraybox}[][c]{l?l}
			\IEEEstrut	
			\eta_c \frac{1 - \frac{ \hat H_c p_{\mathsf{s},c} }{\beta + \hat H_c p_{\mathsf{s},c}}}{\beta}p_{\mathsf{s},c} e^{-\frac{p_{\mathsf{s},c}(h-\hat H_c)}{\beta}}, &\textrm{if $h>\hat H_c$} ; \\
			\eta_c \frac{1 - \frac{ \hat H_c p_{\mathsf{s},c} }{\beta + \hat H_c p_{\mathsf{s},c}}}{\beta}p_{\mathsf{s},c} ,&\textrm{if $0 \le h \le \hat H_c$}, \nonumber
			\IEEEstrut
		\end{IEEEeqnarraybox}
		\right. 	
	\end{iarray}
	wherein $\beta$ is the unique positive solution of 
	\begin{equation}
		\label{42}
		\nu(\beta) \triangleq \beta + \sum_{c=1}^C \frac{ \eta_c \hat H_c p_{\mathsf{s},c} }{\beta + \hat H_c p_{\mathsf{s},c}}-1=0,
	\end{equation}
	with $\sum_{c=1}^C \frac{\eta_c}{\hat H_c p_{\mathsf{s},c}}>1$.
\end{theorem}
\begin{IEEEproof}
First, we will show that the system enters the stage wherein $y=t-h>0$ eventually with probability one. Denote $\hat H_\mathsf{m}=\max_c \{\hat H_c\}$ and $ p_\mathsf{m} = \min_c \{p_{\mathsf{s},c}\}$. Then for $t>\hat H_\mathsf{m}$, every agent successfully transmits an update with a rate no less than $p_\mathsf{m}$. Denote the first time agent-$i$ transmits an update successfully as $t_{i,0}$, then $\forall T>0$,
\begin{equation}
	\Pr\left(\hat t_{i,0} \le T- \hat H_\mathsf{m}\right) \ge  1-e^{-p_{\mathsf{m}}\left(T-\hat H_\mathsf{m}\right)},
\end{equation}
which approaches $1$ with large $T$. After $\hat t_{i,0}$, AoI returns to zero and we have $t>h$ for every $t > \hat t_{i,0}$ afterwards. 

Therefore, with probability one, the solution of \eqref{pdeG} converges to one given by the initial condition \eqref{init2} with $y>0$. Solving that gives
\begin{iarray}
	\hat g_{c}(x,y)  & = & \left\{\,
	\begin{IEEEeqnarraybox}[][c]{l?l}
		\IEEEstrut	
		\kappa_c e^{-\frac{p_{\mathsf{s},c}(x-\hat H_c)}{1-\sum_{j=1}^C \hat G_{j}(\hat H_j,y)}}, &\textrm{if $h>\hat H_c$} ; \\
		\kappa_c ,&\textrm{if $0 \le h \le \hat H_c$},
		\IEEEstrut
	\end{IEEEeqnarraybox}
	\right. 	
\end{iarray}
where $\kappa_c=\frac{\eta_c- \hat G_{c}(\hat H_c,y)}{1-\sum_{j=1}^C \hat G_{j}(\hat H_j,y)}p_{\mathsf{s},c}$, $\forall c$. Note that 
\begin{equation}
	\label{kappa}
	\hat G_{c}(\hat H_c,y) = \kappa_c \hat H_c,\forall c,\,\textrm{and }\sum_{c=1}^C \frac{\kappa_c}{p_{\mathsf{s},c}} = 1.
\end{equation}
Denote $\beta=1-\sum_{j=1}^C \hat G_{j}(\hat H_j,y)>0$ as the total fraction of agents with AoI larger than thresholds, then
\begin{equation}
	\hat G_{c}(\hat H_c,y)  = \frac{ \eta_c \hat H_c p_{\mathsf{s},c} }{\beta + \hat H_c p_{\mathsf{s},c}},\forall c,\textrm{ and \eqref{42}}.
\end{equation}
Observing \eqref{42}, it follows that $\nu(0)=0$, $\nu(\infty)=\infty$, and 
\begin{equation}
	\nu^\prime(\beta) \triangleq 1- \sum_{c=1}^C \frac{ \eta_c \hat H_c p_{\mathsf{s},c} }{\left(\beta + \hat H_c p_{\mathsf{s},c}\right)^2},\, \nu^{\prime \prime}(\beta)>0.
\end{equation}
Hence as long as $\nu^\prime(0) = 1- \sum_{c=1}^C \frac{\eta_c}{\hat H_c p_{\mathsf{s},c}}$ is strictly negative, there is a unique positive solution of \eqref{42}; otherwise there is no positive solution. This concludes the proof.
\end{IEEEproof}

The equilibrium also satisfies the traditional equation
\begin{equation}
\frac{\partial{\hat f_{c}(t,h)}}{\partial{t}}  =0 ,
\end{equation}
which gives the exact same solution as in \eqref{stas}. It is not difficult to check the local-stability of the solution.
\begin{prop}
	For every $\epsilon>0$, there exists $\delta > 0$, such that, if $\|\hat f_{c,0}(h) - \hat d_{c}(h)\| < \delta$, then for every $t\ge0$, $\|\hat f_{c,t}(h) - \hat d_{c}(h)\| < \epsilon$, i.e., the equilibrium $\hat d_{c}(h)$ is asymptotically stable in the local sense.
\end{prop}
\begin{IEEEproof}
	Let us check the Jacobian matrix $\J(\hat d_{c}(h)) \in \mathbb{H} \times \mathbb{H}$, where $\mathbb{H}$ is a Hilbert space with inner product defined as $\langle{f,g}\rangle = \int_{\hat \S} f(x)g(x)\dx x$. In this case $\J(\hat d_{c}(h))$is a the derivative of the right-hand side of \eqref{pde} with respect to $\hat f_{c}(t,h)$, taken on the value of $\hat d_{c}(h)$:
	\begin{iarray}
	&& \J(\hat d_{c}(h))  =  \left\{\,
	\begin{IEEEeqnarraybox}[][c]{l?l}
		\IEEEstrut	
		- \frac{p_{\mathsf{s},c}}{1-\sum_{c=1}^C \hat F_{j}(t,\hat H_j)} , &\textrm{if $0 \le \hat d_{c}(h) < \kappa_c$} ; \\
		0 ,&\textrm{if $\hat d_{c}(h) \ge \kappa_c$},
		\IEEEstrut
	\end{IEEEeqnarraybox}
	\right. 	
\end{iarray}
with the only eigenvalue being $- \frac{p_{\mathsf{s},c}}{1-\sum_{c=1}^C \hat F_{j}(t,\hat H_j)}\kappa_c$ which is strictly negative. Therefore, we can conclude that the equilibrium is locally asymptotically stable based on the Hartman–Grobman Theorem \cite{hg07}.
\end{IEEEproof}
\section{Application 1: AoI in Heterogeneous I.I.D. Multiaccess Channels}
Next, we apply the fluid limit tool to solve a well-known, but open AoI scheduling problem.
\label{sec_app1}
\subsection{System Model}
Consider a one-hop wireless network wherein a central node communications with $N$ distributed agents. The agents share the wireless channel based on a scheduling policy denoted by $\boldsymbol{\pi}$. First assume a time-slotted status update system is considered. Later, we will analyze it using the fluid limits which exist in the continuous time domain. The status packet generation is assumed to be generate-at-will, i.e., a fresh status for agent-$n$ is generated whenever it is scheduled. We are interested in average AoI. Concretely, the long time-average AoI of the system is defined by
\begin{equation}
\label{AoI}
\bar{\Delta}_{\boldsymbol{\pi}} \triangleq \limsup_{T \to \infty} \frac{1}{T N}\sum_{t=1}^T \sum_{c=1}^C \sum_{n=1}^{N_c} \mathbb{E}[h_{n,\boldsymbol{\pi}}(t)],
\end{equation}
where $T$ is the time horizon length, and $h_{n,\boldsymbol{\pi}}(t)$ denotes the AoI of terminal-$n$ at the $t$-th time slot under policy $\boldsymbol{\pi}$. The $c$-th class of agents are those who transmit with a channel reliability of $p_{\mathsf{s},c} \in (0,1]$, $c=1,\cdots.C$. which denotes the successful delivery probability. In each time slot, only one agent can transmit. Over time, the AoI increases by one every time slot.

Such a system model has been studied extensively in the literature \cite{kadota18,ali20,jiang19_tcom,hsu18,talak_mobihoc}. In \cite[Theorem 6]{kadota18}, it is found that there exist a lower bound of the average AoI, i.e.,
\begin{equation}
\label{lb}
\bar{\Delta}_{\boldsymbol{\pi}} \ge \frac{N}{2C^2} \left( \sum_{c=1}^C \frac{1}{\sqrt{p_{\mathsf{s},c}}}\right)^2 \triangleq \bar{\Delta}_{\mathsf{LB}}.\footnote{A slight difference with \cite{kadota18} is that we assume, aligned with the system model in this paper, there are a finite number of $C$ classes of agents and, without loss of generality, each class has $N/C$ agents. In contrast, each agent can be different in \cite{kadota18}, i.e., $C=N$.}
\end{equation}
However, as far as we know, it is still an open problem whether this lower bound is (asymptotically) tight. On the other hand, several Whittle's index based solutions, as well as Lyapunov-drift based ones which essentially gives similar algorithms and performance, have been proposed as a heuristic to solve the problem. These solutions exhibit near-optimal AoI by simulations, but only very loose performance bounds can be proved, e.g., in \cite[Theorem 17]{kadota18}, due to the non-renewal nature of the scheduler. Note that Ref. \cite{ali20} considers a different scenario wherein the number of scheduled agents (i.e., $M$) scales with the total number of agents (i.e., $N$) and goes to infinity with $M/N=\alpha$ fixed. The authors proved asymptotic optimality without specific performance expressions.

\subsection{Fluid Limits Results}
Powered by the results in Section \ref{sec_mf}, we will solve this problem. we will show that the lower bound in \eqref{lb} is indeed asymptotically tight, i.e.,
\begin{equation}
\lim_{N \to \infty} \min_{{\boldsymbol{\pi}}} \frac{\bar{\Delta}_{\boldsymbol{\pi}}}{\bar{\Delta}_{\mathsf{LB}}} = 1.
\end{equation}
In the meantime, we will show that a threshold-based scheduling policy derived from our analysis achieves the lower bound.
\begin{algorithm}
	\caption{Threshold-Based Scheduler 1}
	\label{alg:1}
	\textbf{Initialization}: 
	The threshold of class  $c$ ($c=1,\cdots,C$) is $H_c = \frac{N \sum_{j=1}^C \frac{1}{\sqrt{p_{\mathsf{s},j}}}}{C\sqrt{p_{\mathsf{s},c}}}$.\\
	\For{$t = 1:T$}{
		Randomly schedule an agent whose AoI is greater than its corresponding threshold with equal probabilities at every time.
	}
\end{algorithm}

Consider the scheduling policy in Algorithm \ref{alg:1}, the following theorem is true. 
\begin{theorem}
	\label{thm_1}
	Algorithm \ref{alg:1} achieves a long time-average AoI of $\bar{\Delta}_{\mathsf{Alg.1}}$, and
	\begin{equation}
	\label{aoi1}
	\bar{\Delta}_{\mathsf{Alg.1}}= \frac{N}{2C^2} \left( \sum_{c=1}^C \frac{1}{\sqrt{p_{\mathsf{s},c}}}\right)^2 + \mathcal{O}(1).
	\end{equation}
\end{theorem}
\begin{IEEEproof}
We apply Theorem \ref{thm_mf} to this application. Because we are interested in the time-average AoI, it is sufficient to consider the stationary regime. With \eqref{stas} and notice $\hat d_{c}(\hat H_c) = \hat H_c \kappa_c < \eta_c$, the average AoI of class-$c$ is hence
\begin{iarray}
\overline {\hat h_c} &=& \int_{0}^\infty \hat d_c(h)h\dx h \nonumber\\
&=& \kappa_c \int_0^{\hat H_c}h\dx h + \int_{\hat H_c}^\infty \kappa_c e^{-\frac{p_{\mathsf{s},c}(h-\hat H_c)}{\beta}} h \dx h \nonumber\\
&=& \frac{1}{2} \kappa_c \hat H_c^2 - \eta_c \hat H_c + \frac{\eta_c^2}{\kappa_c} > \frac{\eta_c^2}{2\kappa_c},
\end{iarray}
where the right-hand-side of the inequality can be approached with $\hat H_c^* \to \frac{\eta_c}{\kappa_c}$. In other words, for any $\varepsilon>0$, let $\hat H_c^* = \frac{\eta_c-\varepsilon}{\kappa_c}$, then $\overline {\hat h_c} = \frac{\eta_c^2+\varepsilon^2}{2\kappa_c}$. Note that with \eqref{kappa}, the average AoI across all classes satisfies
\begin{iarray}
\overline {\hat h} &\ge& \sum_{c=1}^C \frac{\eta_c^2 + \varepsilon^2}{2\kappa_c}=\sum_{c=1}^C \frac{\eta_c^2+ \varepsilon^2}{2\kappa_c}\sum_{c=1}^C \frac{\kappa_c}{p_{\mathsf{s},c}} \nonumber\\
&\ge& \frac{1}{2} \left( \sum_{c=1}^C \sqrt{\frac{\eta_c^2+ \varepsilon^2}{p_{\mathsf{s},c}}}\right)^2.
\end{iarray}
The last inequality is based on the Cauchy-Schwarz Inequality, and the equality holds when $\frac{\kappa_c^2}{(\eta_c^2 + \varepsilon^2) p_{\mathsf{s},c}} = C_0$, $\forall c$. Together with $\hat H_c^* = \frac{\eta_c-\varepsilon}{\kappa_c}$, we solve for the following problem:
\begin{equation}
\hat{\H}^* = \argmin_{\hat{\H}_c} \overline{\hat h},
\end{equation}
with the solution of
\begin{equation}
\label{thOpt}
\hat{\H}^* = \left(\frac{\eta_c-\varepsilon}{\sqrt{(\eta_j^2+\varepsilon^2)p_{\mathsf{s},c}}}\sum_{j=1}^C \sqrt{\frac{\eta_j^2+\varepsilon^2}{p_{\mathsf{s},j}}}\right)_{c=1,\cdots,C}.
\end{equation}
Now let us check the legitimacy of the solution against the requirement $\sum_{c=1}^C \frac{\eta_c}{\hat H_c p_{\mathsf{s},c}}>1$ in Theorem \ref{thm_sta}. Indeed,
\begin{equation}
	\sum_{c=1}^C \frac{\eta_c}{\hat H_c p_{\mathsf{s},c}} = \frac{\sum_{c=1}^C  \frac{\eta_c}{\eta_c - \varepsilon} \sqrt{\frac{\eta_c^2+\varepsilon^2}{p_{\mathsf{s},c}}} }{\sum_{j=1}^C \sqrt{\frac{\eta_j^2+\varepsilon^2}{p_{\mathsf{s},j}}}}>1.
\end{equation}
Because $\varepsilon$ can be arbitrarily small\footnote{In the sequel, we omit $\varepsilon$ for brevity since it does not affect the result.} and the fluid limits bring in an approximation error of $\mathcal{O}\left(\frac{1}{N}\right)$, the AoI of Algorithm \ref{alg:1} is 
\begin{equation}
\bar{\Delta}_{\mathsf{Alg.1}}^* = \frac{N}{2} \left( \sum_{c=1}^C \frac{\eta_c}{\sqrt{p_{\mathsf{s},c}}}\right)^2 + \mathcal{O}(1)
\end{equation}
With $\eta_c = 1/C$, we obtain the thresholds in Algorithm \ref{alg:1} and its average AoI, which concludes the proof.
\end{IEEEproof}
\begin{coro}
	Algorithm \ref{alg:1} is asymptotically optimal.
\end{coro}
\begin{IEEEproof}
	Taking the limit $N \to \infty$ with \eqref{aoi1} and \eqref{lb}, the conclusion follows immediately.
\end{IEEEproof}
\begin{remark}
	We would like to draw a connection of Algorithm \ref{alg:1} and the index policies in \cite{kadota18}. The Whittle's index policy and Lyapunov drift-based max-weight policy share the same structure. They both schedule the agent with the maximum value of $p_{\mathsf{s},c} h^2_{c,i}(t)$, with a slight difference in the linear term of $h_{c,i}(t)$ which is insignificant when AoI is large. Essentially, since only relative value is important, this is equivalent to an index of approximately $\sqrt{p_{\mathsf{s},c}} h_{c,i}(t)$. Furthermore, this is analogues to having a clock for agent-$i$ that is scaled by $\sqrt{p_{\mathsf{s},c}}$. In comparison, the AoI threshold of agent-$i$ in Algorithm \ref{alg:1} is $\frac{1}{C \sqrt{p_{\mathsf{s},c}}}\sum_{j=1}^C \frac{1}{\sqrt{p_{\mathsf{s},j}}}$, which in fact scales with AoI by the same coefficient. After all, only agents with AoI larger than the threshold are eligible for being scheduled. Hence, the connection between Algorithm \ref{alg:1} and index policies is very strong.
\end{remark}
\begin{remark}[Decentralized Implementation]
	One additional merit of Algorithm \ref{alg:1}, compared with index policies, is that it is convenient for decentralized protocols. Index policies need to compare the indices of all agents which are difficult to implement in a decentralized system setting. Fortunately, in Algorithm \ref{alg:1}, the central node only needs to broadcast a set of thresholds which do not change over time, and agents can access the channel based upon the thresholds and a contention-based multi-access protocol such as standard Carrier-Sensing Multiple-Access (CSMA). 
\end{remark}
\begin{figure*}[!t]
	\centering
	\includegraphics[width=1\textwidth]{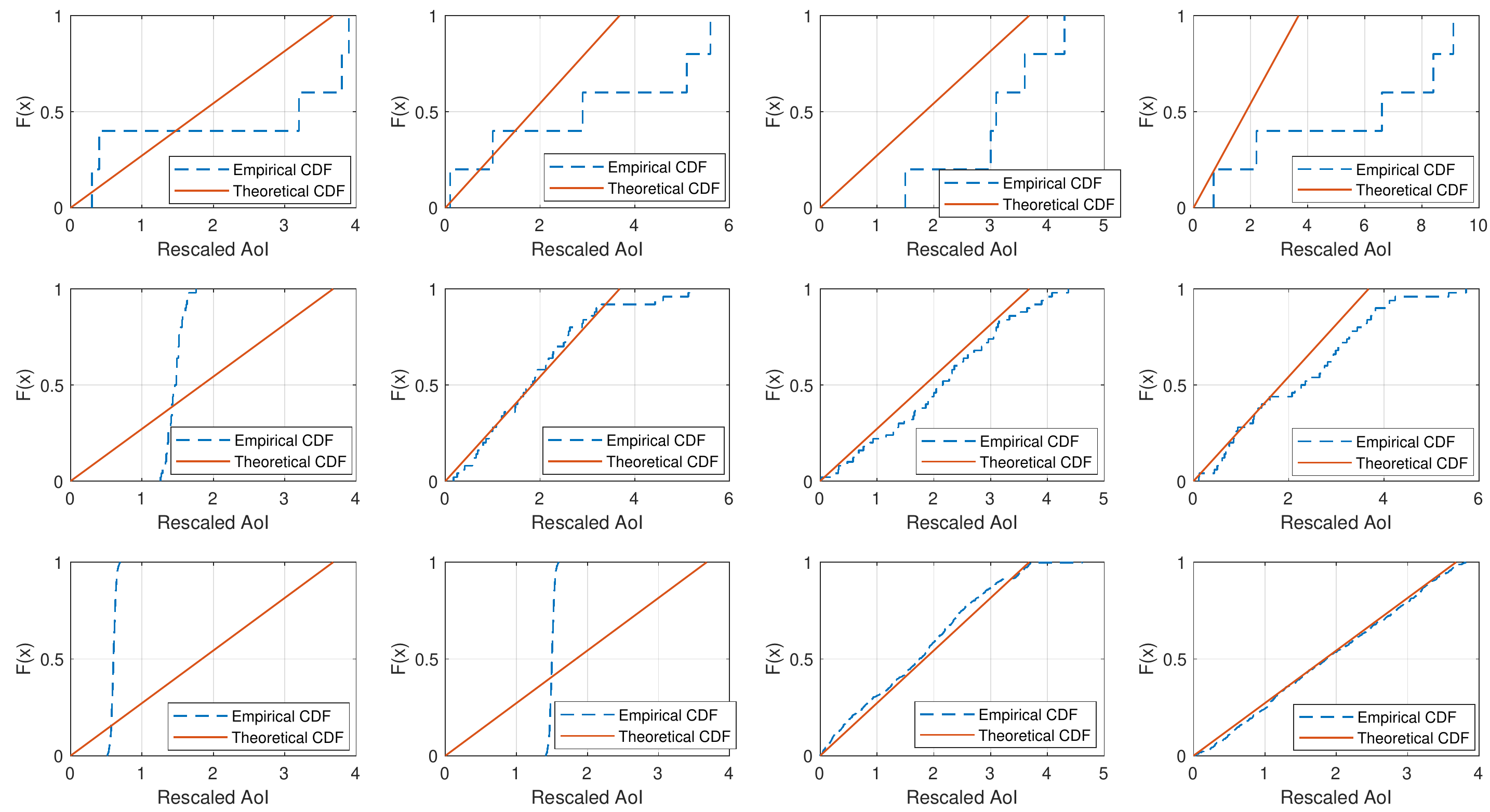}
	\caption{Empirical and theoretical CDF of the rescaled AoI with number of agents of $10$ (top row), $100$ (middle row) and $1000$ (bottom row). From left to right, the four columns represent CDFs at time $100$, $1000$, $10000$ and $50000$ (not rescaled). }
	\label{Fig_aoicdf}
\end{figure*}
\begin{figure}[!t]
	\centering
	\includegraphics[width=0.45\textwidth]{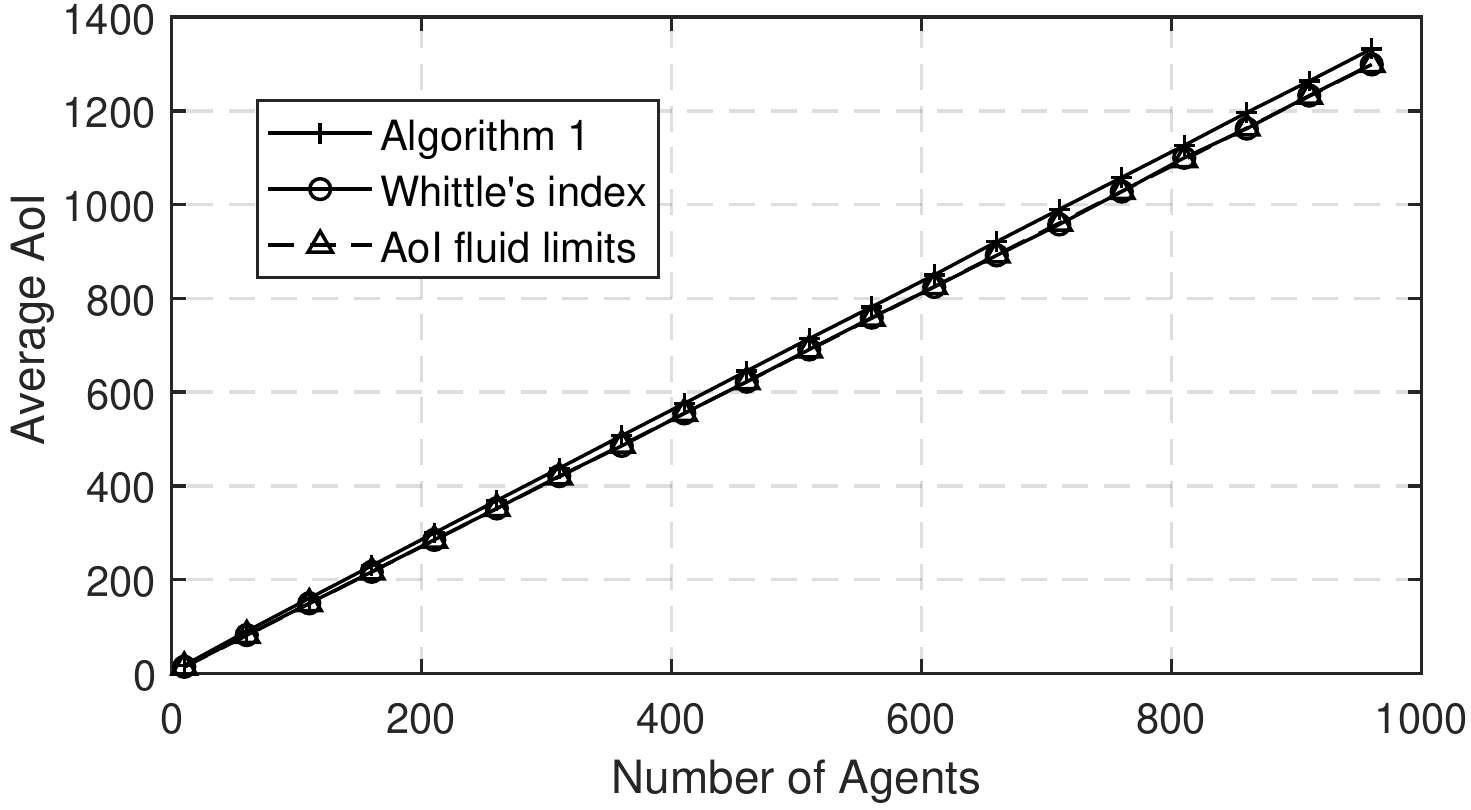}
	\caption{Average AoI comparisons obtained by Algorithm \ref{alg:1}. Whittle's index \cite{kadota18} and fluid limits in Theorem \ref{thm_1}.}
	\label{Fig_aoiavg}
\end{figure}
\subsection{Simulations}
Computer simulations are conducted. Two classes of agents (equal number of agents in each class) are tested, with $p_{\mathsf{s},1}=0.9$ and $p_{\mathsf{s},1}=0.2$ respectively. In Fig. \ref{Fig_aoicdf}, the convergence of empirical CDF of rescaled AoI to the limiting equilibrium of \eqref{stas} based on applying Algorithm \ref{alg:1} is shown. The initial AoI distribution is generated based on a Gaussian distribution with mean of $N/2$ and variance $N$. The simulation is ran in slotted time domain and lasts for $10^6$ time slots. Each slot corresponds to a scheduling event in CTMC. It is observed that when $N$ grows, the match between empirical CDF and theoretical results by fluid limits is evident. Note that it takes longer to converge with larger $N$ because our rescaled AoI requires the time is accelerated by a factor of $N$, which can be seen by comparing the $(2,2)$ and $(3,2)$ figures. We also compare the time-average AoI performance of Algorithm \ref{alg:1}, the Whittle's index approach \cite{kadota18} and the theoretical fluid limits (without the $\mathcal{O}(1)$ term and thus is also the AoI lower bound in \eqref{lb}). It is found that the three are very close to each other. Note that Theorem \ref{thm_1} gives the performance of Algorithm \ref{alg:1} within an error term of $\smallO{(N)}$, but the error can still be dependent on $N$ with lower order than $1$, e.g., $\mathcal{O}(\sqrt{N})$---this explains the widened gap when $N$ grows.

\section{Application 2: Non-Liner AoI Functions}
\label{sec_app2}
In the literature, many works have considered the scenario wherein the value of information is characterized by a non-linear function of AoI \cite{sun19,kosta17,xi19,ayan19}. Let us define the non-linear age function as $V(h)$. Typical forms of $V(h)$ which will be considered in this paper include power and logarithm functions:
\begin{equation}
V_1(\hat h) = \hat h^m,\, V_2(\hat h) = \log{(1+a \hat h)},
\end{equation} 
where $m,a\in \mathbb{R}_{> 0}$ without loss of generality. The non-linear function takes into account the case wherein the value of AoI, reflected on the system performance, grows super- or sub-linearly with AoI, which is common in practice.
\subsection{System Model}
The system setup is similar with Section \ref{sec_app1}, except that instead of minimizing the average AoI, we are trying to minimize a function of AoI. The agents are still classified into $C$ classes. In each class, agents have the same $p_{\mathsf{s},c}$. We assume all agents share the same non-linear age function\footnote{It is not difficult to generalize to the case wherein agents of different classes have different age functions, which would make the denotations much heavier and results less tractable, in which case numerical solutions are needed.}.
\subsection{Fluid Limits Results}
Following the same methodology and algorithm design in Section \ref{sec_app1}, we can prove that the liming PDE exists with the optimal thresholds to be decided. Using the same denotations in \eqref{stas}, we have
\begin{iarray}
	\label{vbar}
	 \overline{V(\hat h_c)} &=& \int_{0}^\infty \hat d_c(h)V(h)\dx h \nonumber\\
	&=& \kappa_c \int_0^{\hat H_c}V(h)\dx h + \int_{\hat H_c}^\infty \kappa_c e^{-\frac{p_{\mathsf{s},c}(h-\hat H_c)}{1-\sum_{j=1}^C \hat F_{j}(t,H_j)}} V(h) \dx h. \nonumber\\
	\overline{V(\hat h)} &=& \sum_{c=1}^C \overline{V(\hat h_c)}.
\end{iarray}
The following theorems derive results for power and logarithm age functions.
\begin{theorem}
	\label{thmV1}
	Consider $V_1(\hat h) = \hat h^m$, the optimal thresholds to minimize $\overline{V(\hat h)}$
	\begin{equation}
	\hat{\H}_{1}^* = \argmin_{\hat{\H}_c} \overline{V_1(\hat h)}=\left(\frac{1}{{p_{\mathsf{s},c}^{\frac{1}{m+1}}}}\sum_{c=1}^C \frac{\eta_c}{{p_{\mathsf{s},c}^{\frac{m}{m+1}}}}\right)_{c=1,\cdots,C},
	\end{equation}
	and the optimum is 
	\begin{equation}
	\overline{V_1(\hat h)}^* = \frac{1}{m+1} \left( \sum_{c=1}^C \frac{\eta_c}{p_{\mathsf{s},c}^{\frac{m}{m+1}}}\right)^{m+1}.
	\end{equation}
	In the unscaled time domain,
	\begin{equation}
	{\H}_{1}^* =\left(\frac{N}{{p_{\mathsf{s},c}^{\frac{1}{m+1}}}}\sum_{c=1}^C \frac{\eta_c}{{p_{\mathsf{s},c}^{\frac{m}{m+1}}}}\right)_{c=1,\cdots,C},
	\end{equation}
	\begin{equation}
	\label{v1}
	\overline{V_1( h)}^* = \frac{N^m}{m+1} \left( \sum_{c=1}^C \frac{\eta_c}{p_{\mathsf{s},c}^{\frac{m}{m+1}}}\right)^{m+1} + \mathcal{O}(N^{m-1}).
	\end{equation}
\end{theorem}
\begin{IEEEproof}
	Following \eqref{vbar}, we obtain
	\begin{iarray}
		\overline{V_1(\hat h_c)} &\ge& \kappa_c \int_0^{\hat H_c}V_1(h)\dx h = \frac{1}{m+1} \frac{\eta_c^{m+1}}{\kappa_c^m}
	\end{iarray}
	\begin{iarray}
		\overline{V_1(\hat h)}^{\frac{1}{m+1}} &=& \frac{1}{{(m+1)}^{\frac{1}{m+1}}} \left(\sum_{c=1}^C \frac{\eta_c^{m+1}}{\kappa_c^{m}}\right)^{\frac{1}{m+1}} \left(\sum_{c=1}^C \frac{\kappa_c}{p_{\mathsf{s},c}}\right)^{\frac{m}{m+1}} \nonumber\\
		&\ge& \frac{1}{{(m+1)}^{\frac{1}{m+1}}}   \sum_{c=1}^C \frac{\eta_c}{p_{\mathsf{s},c}^{\frac{m}{m+1}}}.
	\end{iarray}
    The last inequality is based on the Hölder's Inequalities. By solving the conditions that equality is upheld we can obtain the optimal threshold as
    \begin{equation}
    \hat{\H}_{1}^* = \left(\frac{1}{{p_{\mathsf{s},c}^{\frac{1}{m+1}}}}\sum_{c=1}^C \frac{\eta_c}{{p_{\mathsf{s},c}^{\frac{m}{m+1}}}}\right)_{c=1,\cdots,C}.
    \end{equation}
    Scaling back to the original unscaled time-domain, since the power function amplifies the error by $N^{m}$, after calculating the fluid limit approximation error of $\mathcal{O}\left(\frac{1}{N}\right)$, the optimum is shown in \eqref{v1}.
\end{IEEEproof}
\begin{theorem}
	Consider $V_2(\hat h) = \log(1+a \hat h)$, the optimal thresholds to minimize $\overline{V_2(\hat h)}$ are the solution $x_c^*(a)$ of the following equations ($a$ is omitted here)
	\begin{iarray}
		\label{kkt2}
		\left\{\,
		\begin{IEEEeqnarraybox}[][c]{l?l}
			\IEEEstrut	
			\log\left(1+x_c^*\right)-x_c^* =  \frac{\lambda}{p_{\mathsf{s},c}}, & \forall c; \\
			\sum_{c=1}^C \frac{\eta_c}{x_c^* p_{\mathsf{s},c}}=\frac{1}{a},\,\hat{H}_{c,2}^* = \frac{1}{a x_c^*} & \forall c,
			\IEEEstrut
		\end{IEEEeqnarraybox}
		\right. 	
	\end{iarray}
	and the optimum is 
	\begin{equation}
	\overline{V_2( \hat h)}^* = \sum_{c=1}^C \eta_c\left(\frac{1}{x_c^*} + \eta_c\right)\log\left( x_c^*+1\right)-\eta_c.
	\end{equation}
	In the unscaled time domain, 
	\begin{equation}
	{\H}_{2}^* = N\hat{\H}_{2}^*,
	\end{equation}
	\begin{iarray}
	\overline{V_2( h)}^* &=& \sum_{c=1}^C \eta_c\left(\frac{1}{x_c^*(Na)} + \eta_c\right)\log\left( x_c^*(Na)+1\right) \nonumber\\
	&& -\eta_c + \mathcal{O}(1).
	\end{iarray}
\end{theorem}
\begin{IEEEproof}
	\begin{iarray}
		\overline{V_2(\hat h_c)} &\ge& \kappa_c \int_0^{\hat H_c}V_2(h)\dx h \nonumber\\
		&=& \left(\frac{1}{a}\kappa_c + \eta_c\right)\log\left(a \frac{\eta_c}{\kappa_c}+1\right)-\eta_c
	\end{iarray}
Consider the problem:
\begin{iarray}
&& \min_{\kappa_c} \sum_{c=1}^C \kappa_c \left(\frac{1}{a}\kappa_c + \eta_c\right)\log\left(a \frac{\eta_c}{\kappa_c}+1\right)-\eta_c \nonumber\\ 
&& \textrm{ s.t., } \sum_{c=1}^C \frac{\kappa_c}{p_{\mathsf{s},c}}=1,\,\kappa_c>0,\forall c
\end{iarray}
One can check the Hessian matrix of the objective function and find out that this is a convex problem. Therefore, we solve for the KKT condition in \eqref{kkt2}, from which we can derive the optimal thresholds.

Consider the ratio between unscaled age function scaled age:
\begin{iarray}
	\left|\frac{\log(1+aN\hat h)}{\hat h}\right| \le aN.
\end{iarray}
Therefore, considering the fluid approximation error of the rescaled age is $\mathcal{O}\left(\frac{1}{N}\right)$, the resultant error on the evaluation on the age function is $\mathcal{O}\left(1\right)$.
\end{IEEEproof}	
\begin{remark}
We can infer from Theorem \ref{thmV1} that the Whittle's index (if indexible) should be approximately $p_{\mathsf{s},c} h^{m+1}_{c,i}(t)$, keeping the highest order term. Intuitively, since the value of information is measured by $V_1(\hat h) = \hat h^m$ which essentially becomes more AoI sensitive when $m$ grows larger than $1$, it is reasonable that the index also puts more emphasis on the age increase, in contrast to the channel conditions. On the other hand, when $0<m<1$, the corresponding scheduler considers more about channel conditions. 
\end{remark}
\begin{figure}[!t]
	\centering
	\includegraphics[width=0.45\textwidth]{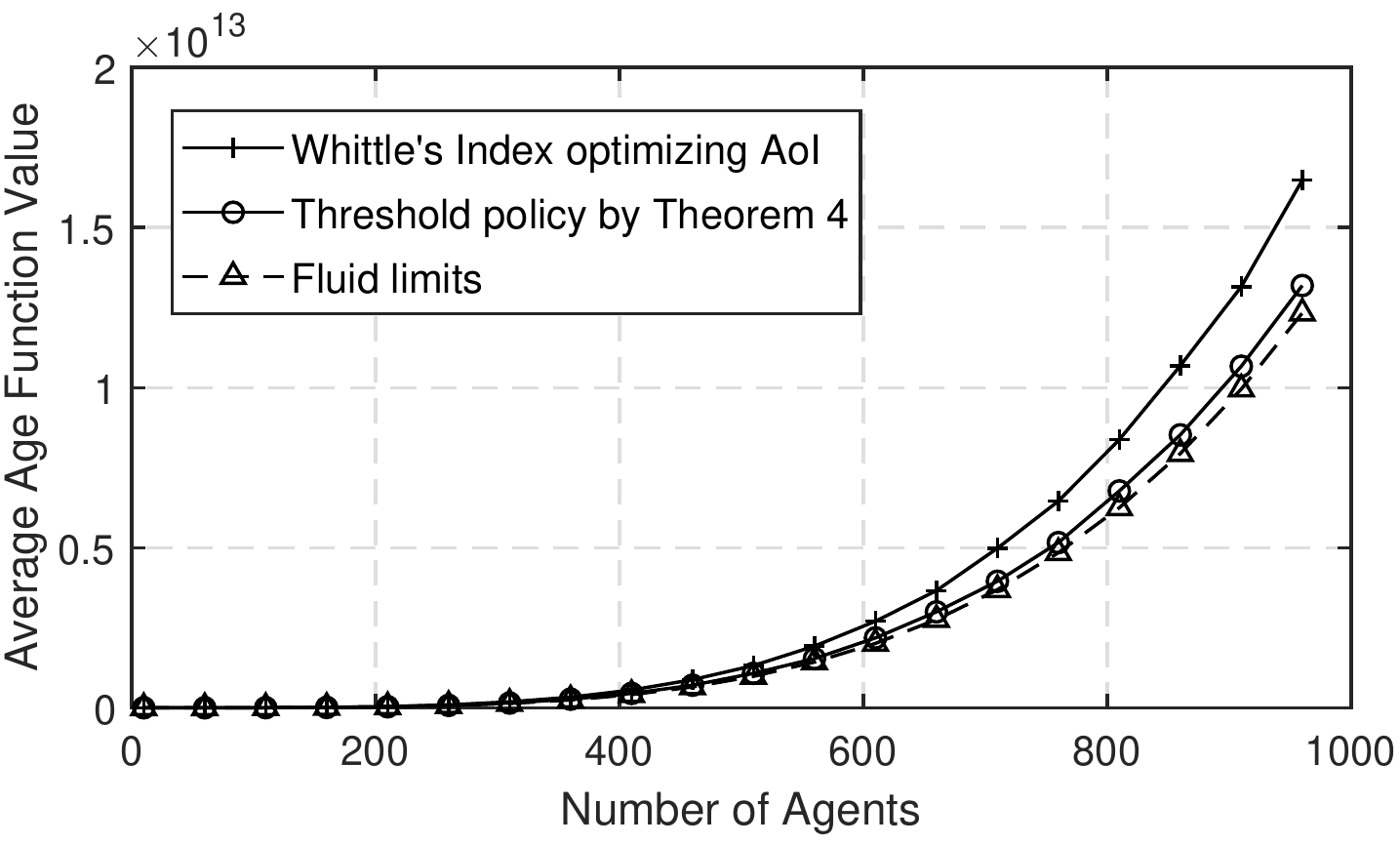}
	\caption{Average non-linear AoI function comparisons obtained by the proposed threshold-based policy, Whittle's index \cite{kadota18} and fluid limits in Theorem \ref{thmV1}.}
	\label{Fig_aoiV1}
\end{figure}
\subsection{Simulations}
In this simulation result, we show in Fig. \ref{Fig_aoiV1} the performance of the proposed threshold-based policy based on Theorem \ref{thmV1}, versus the AoI index policy without considering the non-linear age functions, and the fluid limits. Two classes of agents (equal number of agents in each class) are tested, with $p_{\mathsf{s},1}=0.9$ and $p_{\mathsf{s},1}=0.1$ respectively. The non-linear function is $V_1(\hat h) = \hat h^4$. It is observed that the threshold policy also matches the fluid limits, and it outperforms AoI-optimized policy significantly.

\section{Conclusions and Outlook}
\label{sec_cl}
In this paper, the system asymptotic behavior with a large number of agents, i.e., the fluid limits, is developed for AoI occupancy measure in wireless multiaccess networks. It is shown that, under a simple threshold-based policy, the system limiting behavior converges to a fluid limit within an error inversely proportional to the number of agents. Moreover, the fluid limit has a provable asymptotically local stable equilibrium that can be used to solve for the stationary AoI distribution. 

Leveraging this framework, we solve an AoI scheduling problem with heterogeneous i.i.d. user channel conditions. A well-known existing lower bound, yet with unknown tightness, is shown to be asymptotically tight. The achievability is proved by optimizing the thresholds in our proposed threshold-based policy. The resultant scheduling policy is asymptotically optimal with explicitly derived thresholds, and is much easier to decentralize compared with the index-based policy. Users in the proposed policy only need to know its fixed individual threshold, whereas users have to compare their indices every time by the index policy.

Furthermore, we show that the non-linear age function can be solved by the fluid limit framework. We derive the time-average and optimal threshold-based policies for both power and logarithm age functions in closed-form. Simulations reveal that our analysis matches the reality closely, and that the convergence to the fluid limit is evident with a moderate number of users.

Essentially, we believe a large scope of AoI scheduling problems can be solved leveraging this powerful analysis tool. The crux is the convergence of AoI occupancy measure to the fluid limit. As long as this is upheld, which is often the case in AoI scheduling, the rest follows conveniently. Generalization to other applications is our ongoing work.


\bibliographystyle{ieeetr}
\bibliography{aoi}

\begin{thebibliography}{10}

\bibitem{kaul12}
S.~Kaul, R.~Yates, and M.~Gruteser, ``Real-time status: How often should one
  update?,'' in {\em IEEE Conf. Comput. Commun. (INFOCOM)}, pp.~2731--2735, Mar
  2012.

\bibitem{li15}
B.~{Li}, R.~{Li}, and A.~{Eryilmaz}, ``Throughput-optimal scheduling design
  with regular service guarantees in wireless networks,'' {\em IEEE/ACM Trans.
  Netw.}, vol.~23, no.~5, pp.~1542--1552, 2015.

\bibitem{kadota18}
I.~Kadota, A.~Sinha, E.~Uysal-Biyikoglu, R.~Singh, and E.~Modiano, ``Scheduling
  policies for minimizing age of information in broadcast wireless networks,''
  {\em IEEE/ACM Trans. Netw.}, vol.~26, pp.~2637--2650, Dec. 2018.

\bibitem{jiang18_itc}
Z.~{Jiang}, B.~{Krishnamachari}, S.~{Zhou}, and Z.~{Niu}, ``Can decentralized
  status update achieve universally near-optimal age-of-information in wireless
  multiaccess channels?,'' in {\em International Teletraffic Congress (ITC
  30)}, vol.~01, pp.~144--152, Sep. 2018.

\bibitem{hsu18}
Y.~{Hsu}, ``Age of information: Whittle index for scheduling stochastic
  arrivals,'' in {\em IEEE Int'l Symp. Info. Theory}, pp.~2634--2638, Jun.
  2018.

\bibitem{ali20}
A.~Maatouk, S.~Kriouile, M.~Assaad, and A.~Ephremides, ``On the optimality of
  the {Whittle's} index policy for minimizing the age of information,'' {\em
  arXiv preprint arXiv:2001.03096}, 2020.

\bibitem{talak_mobihoc}
R.~Talak, S.~Karaman, and E.~Modiano, ``Optimizing information freshness in
  wireless networks under general interference constraints,'' in {\em ACM Int.
  Symp. Mobile Ad Hoc Netw. Comput. (MobiHoc)}, pp.~61--70, 2018.

\bibitem{jiang18_iot}
Z.~{Jiang}, B.~{Krishnamachari}, X.~{Zheng}, S.~{Zhou}, and Z.~{Niu}, ``Timely
  status update in wireless uplinks: Analytical solutions with asymptotic
  optimality,'' {\em IEEE Internet of Things Journal}, vol.~6, pp.~3885--3898,
  Apr 2019.

\bibitem{gal12}
R.~G. Gallager, {\em Discrete stochastic processes}, vol.~321.
\newblock Springer Science \& Business Media, 2012.

\bibitem{whitt84}
W.~Whitt, ``Minimizing delays in the {GI/G/1} queue,'' {\em Operations
  Research}, vol.~32, no.~1, pp.~41--51, 1984.

\bibitem{ceran19}
E.~T. {Ceran}, D.~{Gündüz}, and A.~{György}, ``Average age of information
  with hybrid {ARQ} under a resource constraint,'' {\em IEEE Trans. Wireless
  Commun.}, vol.~18, pp.~1900--1913, March 2019.

\bibitem{tang20}
H.~{Tang}, J.~{Wang}, L.~{Song}, and J.~{Song}, ``Minimizing age of information
  with power constraints: Multi-user opportunistic scheduling in multi-state
  time-varying channels,'' {\em IEEE J. Select. Areas Commun.}, vol.~38, no.~5,
  pp.~854--868, 2020.

\bibitem{jiang19_tcom}
J.~{Sun}, Z.~{Jiang}, B.~{Krishnamachari}, S.~{Zhou}, and Z.~{Niu},
  ``Closed-form {W}hittle’s index-enabled random access for timely status
  update,'' {\em IEEE Trans. Commun.}, vol.~68, no.~3, pp.~1538--1551, 2020.

\bibitem{eth09}
S.~N. Ethier and T.~G. Kurtz, {\em Markov processes: characterization and
  convergence}, vol.~282.
\newblock John Wiley \& Sons, 2009.

\bibitem{mftt13}
L.~Bortolussi, J.~Hillston, D.~Latella, and M.~Massink, ``Continuous
  approximation of collective system behaviour: A tutorial,'' {\em Performance
  Evaluation}, vol.~70, no.~5, pp.~317--349, 2013.

\bibitem{kol12}
A.~Kolesnichenko, V.~Senni, A.~Pourranjabar, and A.~Remke, ``Applying
  mean-field approximation to continuous time {Markov} chains,'' in {\em
  International Autumn School on Rigorous Dependability Analysis Using Model
  Checking Techniques for Stochastic Systems}, pp.~242--280, Springer, 2012.

\bibitem{gossip09}
A.~Chaintreau, J.-Y. Le~Boudec, and N.~Ristanovic, ``The age of gossip: spatial
  mean field regime,'' {\em ACM SIGMETRICS Performance Evaluation Review},
  vol.~37, no.~1, pp.~109--120, 2009.

\bibitem{bord07}
C.~Bordenave, D.~McDonald, and A.~Proutiere, ``A particle system in interaction
  with a rapidly varying environment: Mean field limits and applications,''
  {\em arXiv preprint math/0701363}, 2007.

\bibitem{kurtz81}
T.~G. Kurtz, {\em Approximation of population processes}.
\newblock SIAM, 1981.

\bibitem{hg07}
E.~A. Coayla-Teran, S.-E.~A. Mohammed, P.~R.~C. Ruffino, {\em et~al.},
  ``{Hartman-Grobman} theorems along hyperbolic stationary trajectories,'' {\em
  Discrete and Continuous Dynamical Systems}, vol.~17, no.~2, p.~281, 2007.

\bibitem{sun19}
Y.~{Sun} and B.~{Cyr}, ``Sampling for data freshness optimization: Non-linear
  age functions,'' {\em J. Commun. Netw.}, vol.~21, no.~3, pp.~204--219, 2019.

\bibitem{kosta17}
A.~Kosta, N.~Pappas, A.~Ephremides, and V.~Angelakis, ``Age and value of
  information: Non-linear age case,'' in {\em IEEE Int'l Symp. Info. Theory},
  pp.~326--330, Jun 2017.

\bibitem{xi19}
X.~{Zheng}, S.~{Zhou}, Z.~{Jiang}, and Z.~{Niu}, ``Closed-form analysis of
  non-linear age of information in status updates with an energy harvesting
  transmitter,'' {\em IEEE Trans Wireless Commun.}, vol.~18, no.~8,
  pp.~4129--4142, 2019.

\bibitem{ayan19}
O.~Ayan, M.~Vilgelm, M.~Kl\"{u}gel, S.~Hirche, and W.~Kellerer,
  ``Age-of-information vs. value-of-information scheduling for cellular
  networked control systems,'' in {\em ACM/IEEE International Conference on
  Cyber-Physical Systems}, pp.~109--117, 2019.

\end{thebibliography}
\end{document}